\documentclass{aa}
\usepackage{graphicx}

\def\asec{\ifmmode ^{\prime\prime}\else$^{\prime\prime}$\fi}
\def\etal{{et\,al. }}
\def\msun{M$_{\odot}$}

\def\grad{$^\circ$}
\def\it{\sl}
\def\degs{\ifmmode ^{\circ}\else$^{\circ}$\fi}
\def\amin{\ifmmode ^{\prime}\else$^{\prime}$\fi}
\def\asec{\ifmmode ^{\prime\prime}\else$^{\prime\prime}$\fi}
\def\fm{\hbox{$.\!\!^{\rm m}$}}            % Fractions of magnitudes
\def\fd{\hbox{$.\!\!^{\rm d}$}}            % Fractions of days
\def\farcs{\hbox{$.\!\!^{\prime\prime}$}}  % Fractions of arcseconds
\def\fhour{\hbox{$.\!\!^{\rm h}$}} % Fractions of hours
\def\h{$^{\rm h}$}\def\m{$^{\rm m}$}

\def\rxj{RX~\,J1554.2+2721}
\def\degs{\ifmmode ^{\circ}\else$^{\circ}$\fi}
\def\amin{\ifmmode ^{\prime}\else$^{\prime}$\fi}
\def\farcm{\hbox{$.\mkern-4mu^\prime$}}

\begin{document}

%   \thesaurus{06         % A&A Section 6: Form. struct. and evolut. of stars
%              (02.01.2;  % Accretion, accretion disks
%               08.02.1:  % binaries: close
%               08.02.3;  % binaries: general
%               08.09.2:  % Stars: individual:
%               09.18.1:  % (ISM:) reflection nebulae
%               13.25.5)} % X-rays: stars

\title{ The  new cataclysmic variable \rxj\  in the period gap}

   \titlerunning{\rxj\ -- new CV   in the period gap}

\author{G.H. Tovmassian\inst{1}, J. Greiner\inst{2}, S.V. Zharikov\inst{1},
          J. Echevarr\'{\i}a\inst{3}, A. Kniazev \inst{4}}

 \authorrunning{Tovmassian \etal}

 \offprints{G.\,T., gag@astrosen.unam.mx}

   \institute{Observatorio Astron\'{o}mico Nacional,\thanks{use for smail P.O. Box 439027,
San Diego, CA, 92143-9027, USA}
 Instituto de Astronom\'{\i}a, UNAM,  Ensenada, B.C., M\'exico
%\email{gag@astrosen.unam.mx; zhar@astrosen.unam.mx; jer@astroscu.unam.mx}
             \and
Astrophysical Institute Potsdam, An der Sternwarte 16,
14482 Potsdam, Germany
%\email{jgreiner@aip.de}
         \and
Instituto de Astronom\'{\i}a, UNAM, Apartado Postal 70-264, 04510 M\'exico, D.F.,  M\'exico
             \and
Special Astrophysical Observatory of RAS, Nizhnij Arkhyz, Russia, 357147}
%\email{akn@sao.ru}

    \date{Version: \today}
%   \date{Received ?? January 2001; accepted ?? 2001}

\abstract{ We report on the results of a  spectroscopic and
photometric  study of a new  cataclysmic variable,  identified  as
optical counterpart of  the X-ray  source RX\,J1554.2+2721
detected  by ROSAT.   The spectroscopic observations  of the
relatively bright ($\sim$ 16\fm5) object  show systematic  radial
velocity variations  with a  semi-amplitude  of $\sim$140 km/sec.
Besides the clear presence of distinct low and high states  there
are periodic photometric  light  variations  with  an amplitude of
about  0.15 magnitude in the R  band.  The orbital period is
2\fhour753 thus being within the period gap, at its upper border.
The flux  distribution in the spectrum of  the object shows a
substantial contribution  of a M4\,{\sc  v} secondary, and also
bears clear signs  of cyclotron emission. Thus, we  classify the
discovered object as  a new member  of the AM  Her class of
magnetic cataclysmic variables. This classification is  further
supported by the soft X-ray spectrum, the  characteristic profiles
of  the  emission lines,  the tomography map  and the  shape of
the  orbital light curve.   A simple fitting of the spectrum in
the  low and high states suggests a reduced mass transfer rate  in
\rxj\, compared to similar  objects outside the period gap.
} \maketitle

\section{Introduction}

Cataclysmic variables (CVs) are close  binary systems in which
mass is transferred from  a red dwarf  star that fills  its Roche
lobe  onto a white dwarf  (WD).  The AM  Her stars, or  polars,
are a subclass  of CVs containing  a   synchronously  rotating,
accreting,   magnetic  white dwarf. The high (B$\geq$5 MG)
magnetic field of the WD prevents the formation of an accretion
disc and channels  the transferred matter  along the magnetic
lines   directly   onto   the   surface   of   the   WD
(Warner \cite{Warner}). The polars, in contrast to many other  CVs
accreting through a disk, do not produce dwarf nova outbursts, but rather
switch erratically from high luminosity states to low states.
These changes in luminosity are believed to be due to a change of
mass transfer rate. The reason of such changes is not clear,
although irradiation of the secondaries in polars might be one of
the contributing factors \cite{kica}. Alternative possibilities
and an extensive discussion of such behavior in polars has been
presented by \cite{how00}.

 As a  rule, the  orbital  period of  a CV  is a
precisely known  physical parameter. Usually, CVs  have orbital periods
ranging from  80 min  up to $\sim  10$ hours with a highly significant
deficiency (called the {\it  orbital period  gap}) at
$2^{\mathrm h}  < P_{orb} <  3^{\mathrm h}$
(e.g., \cite{Robinson, Ritter}).   Over the
past decade, only $\sim 25$ CV systems were
reported to have periods between 2\fhour0 and
3\fhour0   among  $\sim 300$ CVs with known orbital periods \cite{Downes}.
  The  existence of  the  period gap  is most  commonly
explained in terms of angular momentum loss during evolution. Magnetic
breaking is  considered the  driving force for  the systems  above the
period gap \cite{rap}. It halts at the periods of $\approx$3 hours.
Then the systems drift to  smaller separations without mass transfer and
thus are
faint  and  mostly unobservable  as  a CV.  At periods of  about  2 hours,
other mechanisms, e.g. gravitational waves, become significant
to ignite the mass flow from the donor  star, and the system regains  its
brightness  because accretion resumes (\cite{how} and references therein).
However, polars are found rather often in the period gap
 either narrowing it substantially or casting doubts on its existence at all.
The observed low value of the mass transfer rate for polars  is
sufficient reason for the absence of a period gap in their distribution \cite{wiwu94}.
As  \cite{li} have shown, the magnetic breaking may not
apply to magnetic systems.

\begin{figure}[t]
\includegraphics[width=105mm, bb=28 28 570 350]{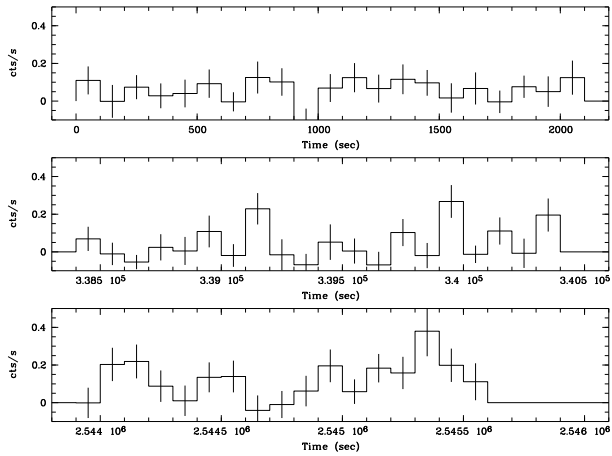}
\caption{X-ray light curve of RX J1554.2+2721 derived from the pointed ROSAT
PSPC observation 800551p in Jul/Aug 1993. The observation was split into
three observation intervals, separated by 3.9 and 25 days, respectively.
Each of the panels shows a  2300 sec stretch of the observation with the
time given relative to the start of the observation.}
\label{rxj1554_xlc}
\end{figure}

Here  we report the discovery of a new cataclysmic variable, called
\rxj\, (=1RXS J155412.7+272143) which  is
of the AM Her type
(the ROSAT source was  very recently identified independently as a CV by
\cite{Jiang}, while this paper was in preparation).
\rxj\, has an orbital period of 2\fhour753, which  places it at the very
edge of the upper limit of  the period gap.
Each new object in the period gap is very important for
theoretical models of the evolution of  CVs.

\section{Observation and data analysis}

\subsection{X-ray observations}

RX  J1554.2+2721 was scanned  during the  ROSAT all-sky-survey  over a
period of 4 days in August 1991 for a total observing time of 380 sec.
Its  mean  count rate  in  the  ROSAT position-sensitive  proportional
counter   (PSPC)   was   0.12    cts/s,   and   the   hardness   ratio
$H\!R1=-0.78\pm0.15$ where $H\!R1$ is  defined as (H--S)/(H+S), with H
(S) being the counts above (below) 0.4 keV over the full PSPC range of
0.1--2.4 keV.

RX J1554.2+2721  was serendipitously covered by a  pointed ROSAT PSPC
observation,  though at  very large  off-axis angle.  This observation
(800551p) was performed in July/August  1993 for 6090 sec, and detects
RX J1554.2+2721  at a (vignetting  corrected) mean count rate  of 0.14
cts/s.  For the  spectral and timing analysis the  source photons were
extracted with a radius of 6\farcm5. The background was chosen between
7.5--10\amin\ wide ring sector  of 270$\degs$ disregarding a 90$\degs$
part along  the rim  of the field  of view. Standard  corrections were
applied using the dedicated  EXSAS software package \cite{Zimmermann}.

\subsubsection{X-ray light curve}

The light curve  of the three separate exposure  intervals of observation
800551p in July/August 1993 is shown in
Fig. \ref{rxj1554_xlc}  with  a  100  sec binning.   Two  things  are
immediately obvious: (i) the count rate varies between a mean floor of
0.1--0.2 cts/s  to up to  0.4 cts/s during  peaks; and (ii)  there are
"dips" of  nearly zero  count rate at  regular intervals of  about 400
sec.  These "dips" are due to  the wobble of the ROSAT satellite which
occasionally  places  RX J1554.2+2721  outside  of the  field-of-view.
Thus, the mean count rate as given above is certainly a lower limit to
the actual  intensity. Apart from this cosmetic effect,  this also makes a
folding over  the best-fit
orbital  period  (see  below)  rather  useless,  since  it  introduces
spurious features in addition to  the low phase coverage caused by the
short exposure time.

\begin{figure}[t]
\includegraphics[width=105mm]{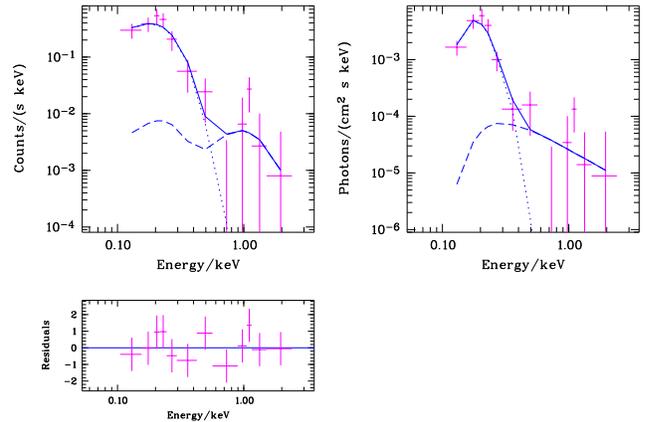}
\caption{ROSAT PSPC X-ray spectrum of RX J1554.2+2721 derived from the pointed
observation 800551p fitted with a sum of a blackbody and a thermal
bremsstrahlung model. The lower left panel shows the deviation between
data and model in units of $\chi^2$ per bin.}
\label{rxj1554_xspec}
\end{figure}

\subsubsection{X-ray spectrum}

The X-ray spectrum  during the pointed observation is very soft
($H\!R1=-0.9\pm0.1$).  Fitting a  one-component model,  e.g.  a simple
black body model, results in a large reduced $\chi^2 = 2.8$. Applying a
sum  of a  black  body and  a  thermal bremsstrahlung  model with  the
temperature  of  the latter  component  fixed to  20  keV  (it is  not
constrained at all  by the ROSAT data) gives a  good reduced $\chi^2 =
0.9$ (see Fig. \ref{rxj1554_xspec})  and the following fit parameters:
$kT_{\rm  bbdy}$ =  17$\pm$15 eV,  $N_{\rm H}$  = 3.3$\times$10$^{20}$
cm$^{-2}$.  For  a better  comparison  to  the  parameters of  similar
sources we also fixed the  black body temperature to $kT_{\rm bbdy}$ =
25  eV,  and  derive  $N_{\rm H}$  =  1.5$\times$10$^{20}$  cm$^{-2}$,
Norm$_{\rm    bbdy}$    =    0.045    and    Norm$_{\rm    thbr}$    =
2.6$\times$10$^{-5}$.  This  gives an unabsorbed 0.1--2.4  keV flux of
2$\times$10$^{-12}$ erg cm$^{-2}$ s$^{-1}$ (or 5$\times$10$^{-12}$ erg
cm$^{-2}$  s$^{-1}$   bolometric),  corresponding  to   an  unabsorbed
bolometric  luminosity   of  6$\times$10$^{30}$  (D/100 pc)$^2$  erg
s$^{-1}$.

\subsection{Optical observations}

\begin{figure}[t]
\includegraphics[width=85mm, bb=100 340 490 720,clip=]{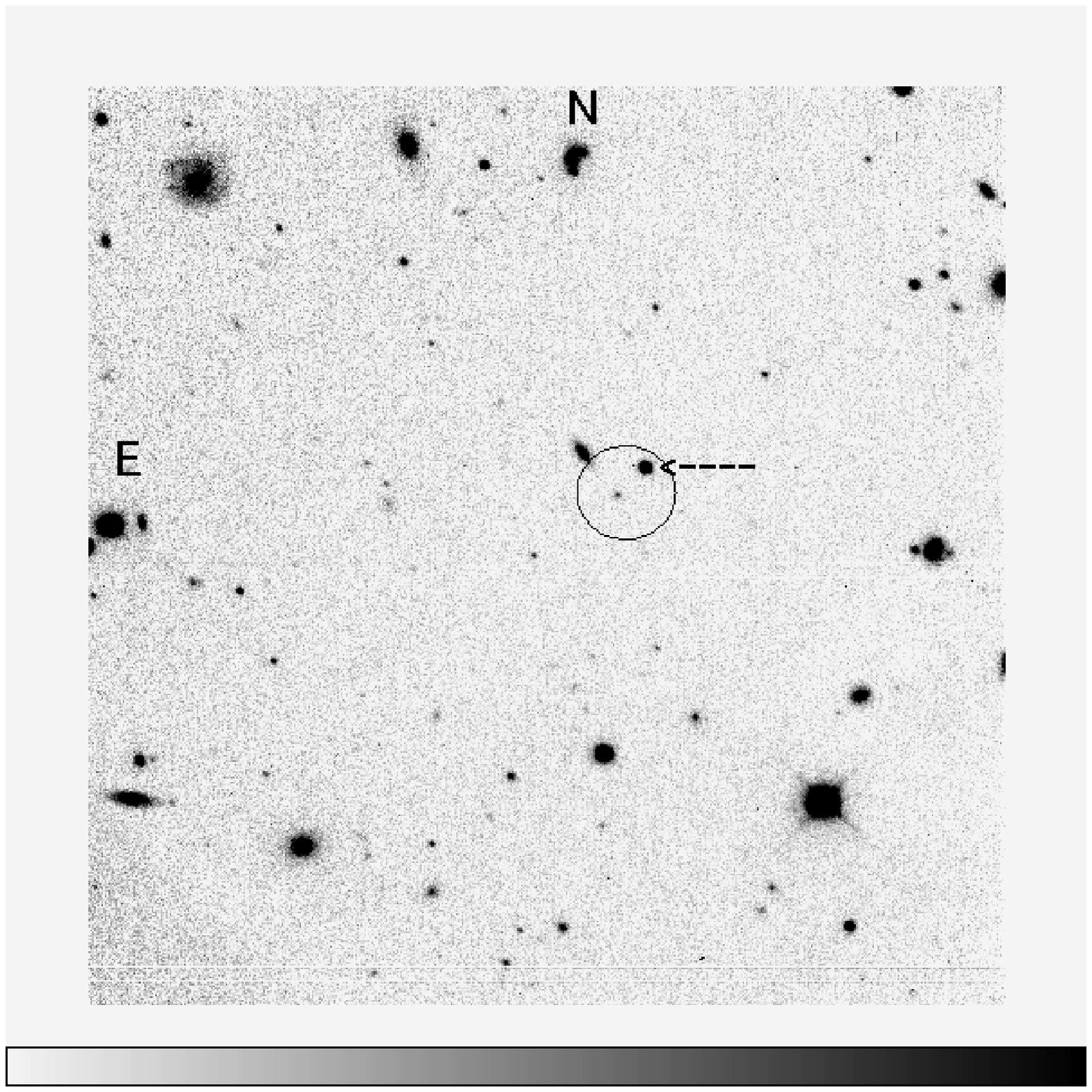}
\caption{Identification chart showing the field of \rxj. The circle in the
 center shows the ROSAT 30\asec\ error box.}
\label{errorbox}
%\vspace{-0.5cm}
\end{figure}

The CCD photometric and spectroscopic observations of  \rxj\ were carried
out at  the 1.5m \&  2.12m telescopes of the  Observatorio Astron\'omico
Nacional de San Pedro M\'artir (OAN SPM), Mexico and  at the 6\,m
telescope of the Special Astrophysical Observatory  of the Russian
Academy of Sciences (SAO RAS).
The  log of observations
is presented in the Table \ref{logObs}.  Inspection of the spectrum of
the brightest object in the ROSAT 30$^{\prime\prime}$ error box
(Fig. \ref{errorbox}) showed
that it has spectral characteristics  typical to a CV.
The USNO catalog gives the following measurements:
   U1125\,\,07424531:  15\h 54\m 12\farcs40 +27\degs 21\amin 51\farcs2,
  R=16\fm5 and B=15\fm9 for this object.

An identification chart  of \rxj\, is given  in Fig. \ref{errorbox}
with the CV marked.

\subsubsection{Optical photometry}

We obtained  time-resolved $R_c$-band photometry of  \rxj\, during
two nights  on May 26 \& August  1st, 2000 at the  1.5m telescope.
Several additional  measurements were obtained in March and October 2000.
The telescope was equipped with a SITE CCD in a direct imaging mode.  In total,
the object  was monitored during $\sim$ 6\fhour3  (3\fhour3 the first
night and 3$^h$ the second night).  The exposure times were 300 sec and 180
sec, respectively, which  leads to a time resolution of  350 \& 230 sec,
taking  into  account the  dead   time  due to  read-outs.   The  standard
reduction techniques of CCD photometry were applied using the DAOPHOT package
in IRAF to measure differential magnitudes.  The images were corrected
for  bias and flat-field  before aperture  photometry was  carried out.
The  dispersion  of magnitudes  ranged  from  0.005  to 0.01  mag  as
measured from the differential photometry of comparison stars with a
similar brightness.  Unfortunately, no absolute calibration for  our
photometric data was obtained.

The light curve shows $\approx 0\fm15$  smooth  brightness
variations. The nightly mean magnitudes fall into two categories,
with 0.6 mag difference between them. The shape of the light curves
changes together with the change of the mean magnitude.
There is no notable flickering or short time scale oscillations.  For
the results of the period analysis and the differences in the light curves
as well as  their interpretation we refer to the following sections.

\subsubsection{Optical spectrophotometry}

Two  spectra of  the object  were taken  on April  11, 2000  using the
Boller \&  Chivens spectrograph installed  in the Cassegrain  focus of
the 2.12m  telescope  of OAN SPM.  The  300\,l/mm grating was  used to
cover a wide range of wavelengths spanning from 4000 to 7600\,\AA.

\begin{table*}[t]
\caption{Log of Optical Observations}
\begin{tabular}{lccccl}
\hline\hline
Date &  HJD  start &   Duration  & Exposures  & Band/Wavelength      & Telescope/Equipment   \\
(2000 year)  &2541000+ & (minutes)  & ( sec)          &               &    \\ \hline
16 Mar  & 619.82       &  118      &   300            &   $R_c$       &  1.5m CCD                   \\
11 Apr  & 646.002      &  20       &   600            &  3680-7580    &  2.12m  B\&Ch spectrograph   \\
12 Apr  & 647.901      &  167      &   600            &  3450-5100    &  2.12m B\&Ch spectrograph     \\
26 May  & 691.748      &  200      &   300            &  $R_c$        &  1.5m CCD      \\
1 Aug   & 758.518      &  180      &   180            &  $R_c$        &  1.5m CCD      \\
18 Sep  & 805.637      &  83       &   720            &  3550-5530    &  2.12m\  B\&Ch spectrograph           \\
05 Oct  & 822.613      &  83       &   720            &  3680-5650    &  2.12m\  B\&Ch spectrograph           \\
06 Oct  & 823.65       &  30       &   600            &  3660-8100    &  6m \ UAGS                          \\
06 Oct  & 823.60       &  10       &   600            &   $R_c$       & 1.5m  CCD       \\
\hline
\end{tabular}
\label{logObs}
\end{table*}

Time-resolved spectroscopy  of the  optical counterpart of  {\rxj} was
obtained on April 12 and September 18, 2000 on the same telescope and same
spectrograph.  We used the 400\,l/mm  grating with a $13^o54$ blaze in
the  second  order,  combined  with  the  blue  BG39  filter  and a
TEK$1024\times 1024$\,pixel CCD with a $0.24\mu$ pixel size.  The slit width
was 1.5 arcsec projected on the sky.  The seeing was satisfactory with
$\leq  1.2$ arcsec.  The exposure times  were 600 and 720
sec.   The spectral range  from $\lambda$\,3800\,\AA\,  to 5100\,\AA\,
was covered  with spectral resolution  of 1.5\,\AA/pix leading  to a
2.8\,\AA\, FWHM resolution. In total, 26 spectra were obtained,
among them eight  in September 2000.

\begin{figure}[t]
\setlength{\unitlength}{1mm}
\resizebox{12cm}{!}{
\begin{picture}(180,140)(0,0)
\put (-10, 0){\includegraphics[width=145mm]{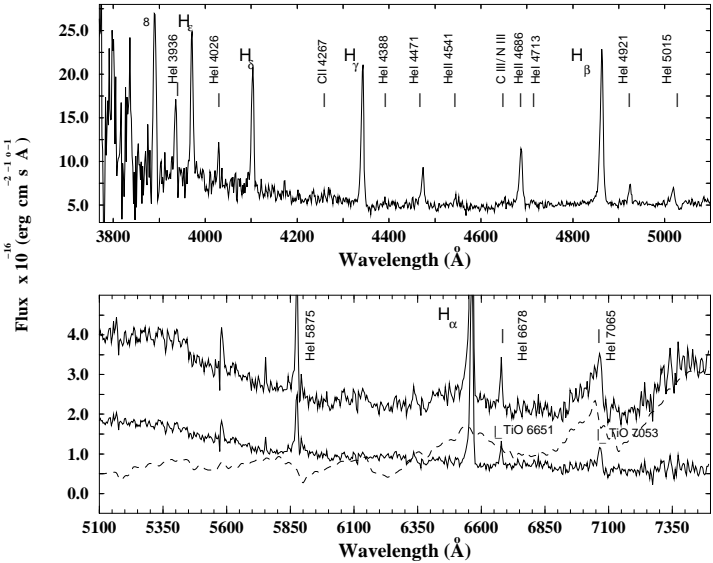}}
%\put (0, 0){\includegraphics[width=250mm]{fig4.eps}}
\end{picture}
}\hfill
\caption{The  sum of high state  spectra of \rxj\ obtained in April 2000.
In the upper  panel higher resolution spectra in the blue part.
In the bottom panel red portion of lower resolution spectra.
Most of the spectral lines are marked.
The spectrum of M4\,{\sc v} standard star and the result of subtraction of the red star
from observed spectrum  of \rxj\ are presented too.}
\label{Spectr}
\end{figure}

A number of additional spectra were obtained in   October 2000.
Two spectra were obtained on October 5 with similar settings
as the previous  September observations at SPM. During the next night
we acquired
three more spectra at the  6\,m telescope of SAO  in Russia, one in the
blue region $\lambda$\,3700--6100\,\AA\,
and two at $\lambda$\,5700--8100\,\AA\, wavelength range,
The UAGS spectrograph with the Photometrix CCD  was used in
order to obtain $\approx$5\,\AA\, FWHM resolution spectra.
He-Ar  arc spectra  were   used during all observations for wavelength
calibration.
In all   cases except the September and October SPM observations,
spectrophotometric  standard stars  were observed  in order  to perform
flux calibration.  The IRAF  long slit spectroscopic reduction package
was used for the extraction  of spectra, wavelength and flux calibrations.
Before these steps, the images were corrected for bias and cosmic rays.

It appeared that the October spectra were obtained when \rxj\, was
in a low state, while the rest of the spectra were acquired with the system
being in a high state.

In  Fig. \ref{Spectr} we present the  average (after correcting the
wavelengths for orbital motion) of all high-resolution spectra and the
red portion  of the low-resolution spectra of \rxj\, in the high
state.  They  exhibit all emission
lines characteristic  to CVs, and  also clear TiO  absorption features from
the M-dwarf secondary.  Some major  spectral features are marked in
Fig. \ref{Spectr}.  A comparison with  spectra of  main-sequence
late type stars yields a spectral type later than M3\,{\sc v} and
earlier than M5\,{\sc v}. A M4\,{\sc v} spectrum is also  presented  in
Fig. \ref{Spectr}.
There is a small  deficiency in  flux at $\lambda$\,7300-7600\,\AA, between
the object and the standard, which may indicate a slightly later type
$\approx$M4.5\,{\sc v}. We subtracted the standard M4\,{\sc v} star spectrum
from the sum  of low-resolution spectra  of \rxj.  The  resulting   spectrum
is   shown   in   the   lower  panel   of  Fig. \ref{Spectr}.

\section{Orbital period}

\begin{figure}[t]
\includegraphics[width=90mm,bb = 10 50 380 370, clip=mm]{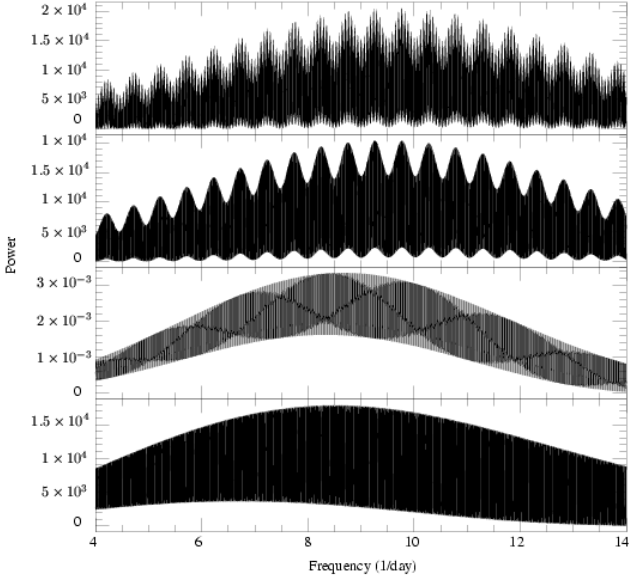}
\caption{The power spectra of \rxj\, obtained from analysis of: (from bottom to top)
         first panel: H$\beta$ NEL radial velocities from April observations;
     second panel: two nights of R band photometry;
     third panel: H$\beta$ RV data with addition of September data;
     forth panel: same as in third panel with addition of October points.
}
\label{powsp}
\end{figure}

The bright H$\beta$ line was deblended with  two Gaussians as a
first step, common for such kind of objects (\cite{sch, tov}). The
resulting components fall into  two  categories according  to
their  widths, amplitudes  of radial velocity (RV)  variations and
phases. The dominant  and thus more easily  distinguishable narrow
emission  line   (NEL)  component shows a FWHM of approximately
4--6\,\AA.  It was separated from the rest, and underwent period
analysis. The power spectrum produces a broad peak around 8.7
day$^{-1}$ if only  the data of one night with the longest
coverage are used. The corresponding power spectrum is presented
in the lower panel of Fig. \ref{powsp}. Its maximum value closely
coincides with  the photometric period. The photometric data also
produce a broad feature in the power spectrum (second panel from
the bottom in Fig. \ref{powsp}). Adding the September data to the
spectroscopic sequence moves the highest peak in the power
spectrum to shorter periods (third panel). Finally, including two
more points obtained in October, shifts the highest  peak in the
power spectrum to a higher frequency ($\approx$9.7\,day$^{-1}$),
as can be seen on the top panel of Fig. \ref{powsp}. However, the
latter value hardly can be acceptable, since the analysis of one
night data does not support the possibility of such short a
period. We selected the maximum peak in the range of values
between 8.5 and 9.0\,day$^{-1}$ as the most plausible for the
orbital period of \rxj. This corresponds to 0\fd11473$\pm$0\fd0006
= 2\fhour7535. This error of period estimate is rather technical,
assuming that we selected the right peak in the power spectrum.
However, we do not have efficient way to distinguish it
unambiguously based on our data. {\sl (when this paper was already
submitted to the publication Thorstensen etal (2001) reported a
2\fhour52 period for this object, which does not contradict
neither our period analysis, nor the main results following from
there)}.
%Although we obtained high precision by combining data separated by
%months, the real orbital period may differ as much as 0\fd0006,
%because we have no effective way to distinguish which of those
%peaks in the power spectra is the most significant.

\begin{figure}[t]
%\setlength{\unitlength}{1mm}
%\resizebox{12cm}{!}{
%\begin{picture}(180,180)(0,0)
\includegraphics[width=90mm]{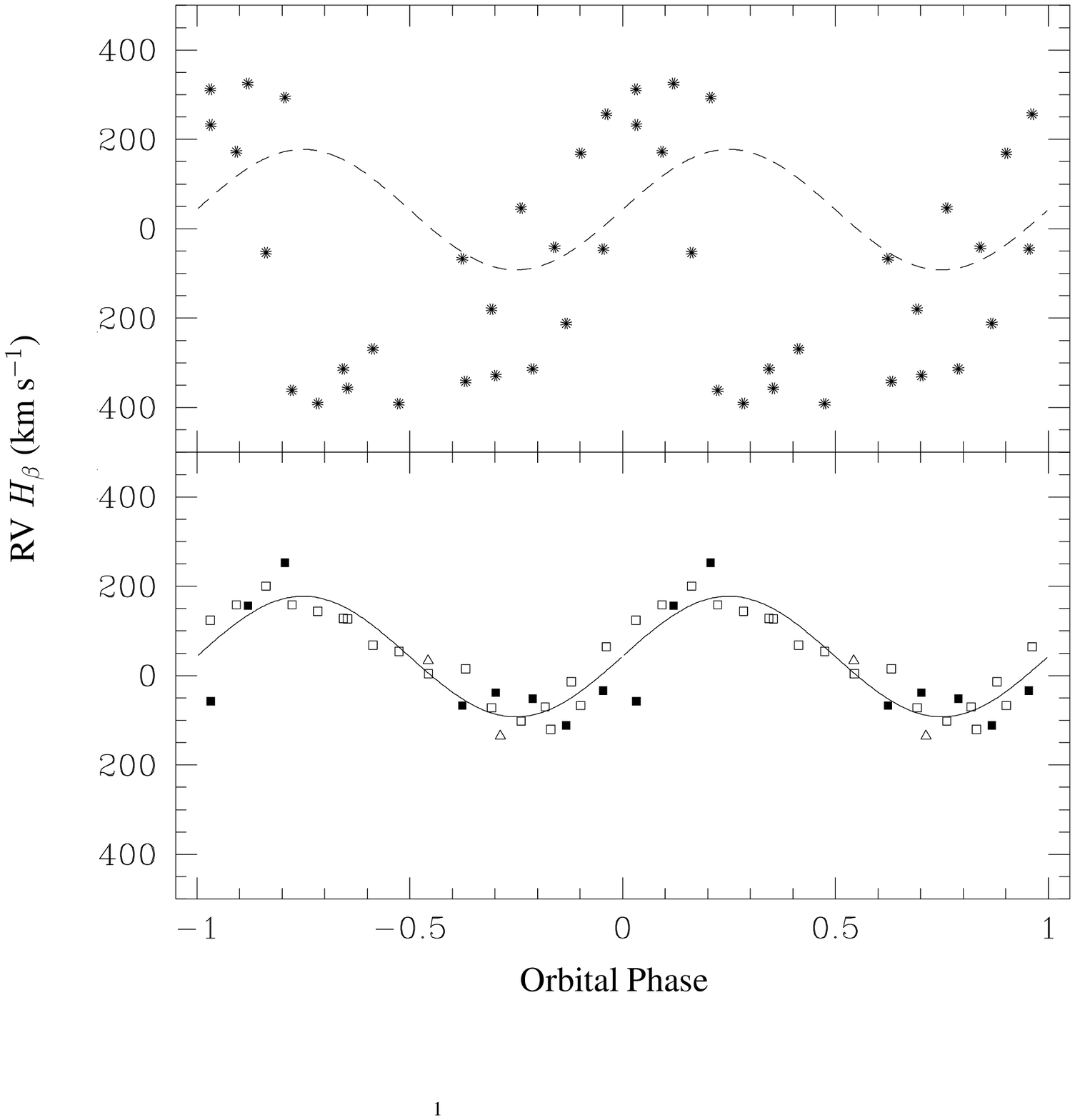}
%\put (0, 0){\includegraphics[width=230mm]{fig6.eps}}
%\end{picture}
  \caption{The  radial velocity curves or the NEL and HV component of
   the H$\beta$  emission line on the bottom and upper panels, respectively.
    The solid line is a $sin$ fit to the NEL radial
   velocities repeated on both panels for comparison.
The open squares mark observations of April, black squares are those from September
and open triangles mark October measurements. By  asterisks in the upper
panel the HVC poorly determined measurements are presented.
}
\label{RadialVel}
\end{figure}

Assuming that
the  NEL component  originates from  the heated  front portion  of the
secondary  star  (as it  is  common in  a  number  of AM\,Her  systems, e.g.
(\cite{sch, tov}), the  radial  velocity  curve  derived from the
NEL measurements  was fitted  with a sine curve with the derived  orbital
period of $P_{orb}=0.11473$ days:
\begin{equation}
v(t) = \gamma_{o}+K_{2}*\sin(2\pi(t-t_o)/P),
\end{equation}
where $\gamma_o$ is  the systemic velocity, and K$_2$
is the semi-amplitude  of the radial velocity of the  secondary star, both
in   km  $s^{-1}$.    The   observation  time   is   $t$,  the   epoch
$t_0=2451761\fd2281\pm0\fd0001$ corresponds  to the -/+  zero crossing
of the  $H\beta$ radial velocity  curve, and therefore is inferior
conjunction   of   the   secondary star.   Our   best   fit   yielded
$K_{NEL}=139$\,km/sec with a standard deviation of $\sigma=43$\,km/sec.
The RV curve and our fit  of the NEL component are presented in the bottom
panel of  Fig. \ref{RadialVel}.

The  other  component,   although
not well distinguishable at some phases, has an amplitude and orbital phasing
typical to the high velocity  component (HVC) as defined in studies of
magnetic CVs (\cite{cro90, sch}). It is assumed to  originate in the ballistic part of the
accretion stream. The HV component of emission lines is shown on the upper
panel of Fig. \ref{RadialVel}. The fit of the radial velocities from the NEL
is over-plotted for comparison.

\section{The system composition: high and low states}
\subsection {Photometry}

As  mentioned above, the  system  switches from  one luminosity
state to the other very frequently. We  present the behavior of
the object in Fig. \ref{highlow}. The  mean differential
magnitudes of the system  in comparison to a nearby star as
measured from photometric observations differ by up to 0\fm65. In the
case of spectral observations, when no flux calibration was
available we can guess the state of the system from the appearance
of the spectrum. It shows that the system undergoes a change of
state  almost every month. Actually, we do not know any other
similar system which does these state changes so frequently.

\begin{figure}[h]
\includegraphics[width=85mm,clip=]{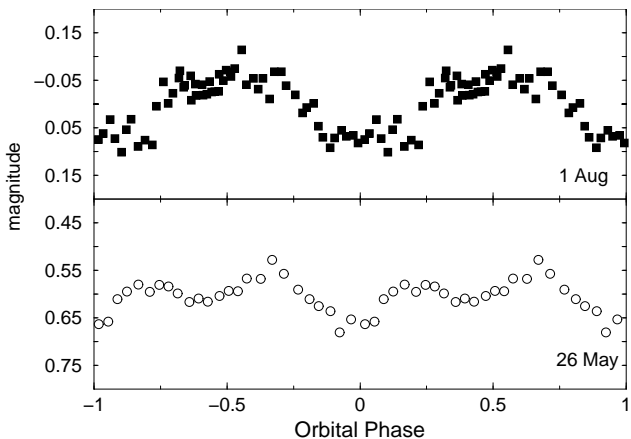}
\caption{Phase folded  $R_c$ light curves of \rxj, separately in
high state (upper panel) and low state (bottom panel).
}
\label{HighLow}
\end{figure}

Fig. \ref{HighLow}  demonstrates  the $R_c$  light curve  of
\rxj\, folded with the orbital  period and phased in accordance to the
above derived orbital parameters.
The bottom panel displays the light curve obtained in the low luminosity state.
The upper panel presents the light curve of \rxj\, while the system was
in the high state. There are several important features which distinguish
these two curves:

\begin{figure}[h]
\includegraphics[width=90mm,clip=]{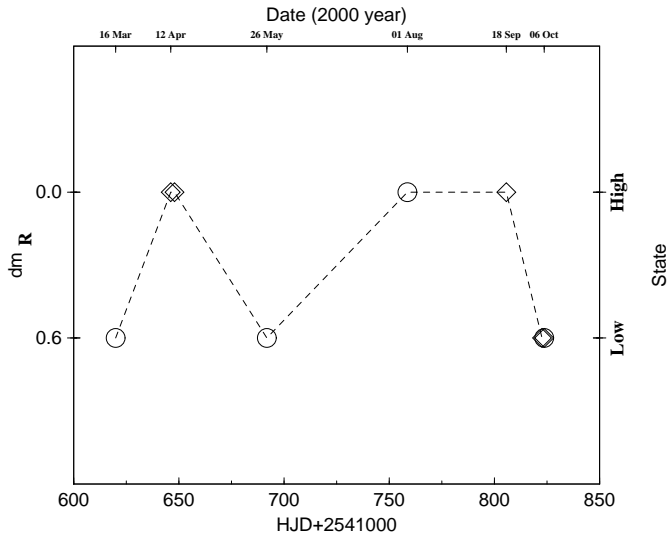}
\caption{The behavior of the system during the  period of our observations.
The circles are differential magnitudes determined from photometry. The left
vertical axes display differential magnitudes in comparison to the field stars.
The rhombs mark spectroscopic determination of the luminosity state and its height is symbolic
marked on the right vertical axes}
\label{highlow}
\end{figure}

In the low state the system shows a pronounced double humped
curve. Taking into account the large contribution of the secondary
in the total light we assume that in the R band during the low
state we see mainly the surface of the secondary star. It is
elliptically distorted by filling its corresponding Roche lobe and
irradiated. This causes the double hump feature with  uneven depth
of the minima. This is due to the changing aspect of the
secondary's Roche-lobe which presents the largest projected area,
and hence highest flux at phases 0.25 \& 0.75. At phase 0.0 \& 0.5
the projected area is the smallest. The flux decrease at phase 0.5
is smaller than at phase 0.0,  because that side of the secondary
has a higher temperature due to illumination, and hence is more luminous
than the back side.

We  estimate the mean amplitude of brightness variation from our
R$_c$ band light curve to be  0.13 magnitude in the low state.
Using the  equation (2.109b) from Warner (\cite{Warner})
we obtain an inclination of the system of $45^o - 55^o$,
assuming a mass ratio  $0.42 < q < 0.62$  for  corresponding  masses of the
white dwarf of $0.39-0.57$\msun\ and secondary mass of
M4\,{\sc v}=0.24\msun\ \cite{schka}.
The equation is very sensitive to a number of parameters.
We were able to confirm and select a plausible solution $q$--$i$ by means
of Doppler tomography (see below).

The light curve in the high luminosity state is  similar, but
the double hump feature is smeared out by some additional emission.
The amplitude of variation is about 0.2 magnitudes, a $\sim 0\fm07$ increase
from the low state value.
The primary minimum is also changed, being broader and flatter
at the bottom.  We suppose that
additional cyclotron emission starts to play a role in the total flux,
when the pole is in view. There is also a significant contribution
of the H$\alpha$ line in the high state in comparison to the low state.
Also, there appears to be some flickering
in the light curve during the high state.

\subsection {Spectrophotometry}

The changes in the spectra of \rxj\, between the high and low states
are even more drastic. The broad band photometry in R includes a large
contribution from the  secondary,
while the change of flux is conditioned by the accretion rate, and thus
significant changes occur in the  blue part of the spectrum.
In Fig. \ref{Photometry} the low resolution spectra of \rxj\,
are presented in both states. The red portion of the spectra
are dominated by the secondary star.
It is very unusual that any cataclysmic variable in a high state,
with a period shorter than 4--6 hours, will
bear clear signs of its secondary in its spectrum at wavelengths shortward of
H$\alpha$.
 Commonly, the
contribution from accretion, either disk in most CVs or magnetic driven in
polars, is the dominating source of light in the whole optical range during
the high state.
Therefore, we conclude that in \rxj\, we have a reduced accretion rate even during the
high state.

Another indication of low accretion  is the narrowness of the emission lines
of the system when it is in the low state. Actually, their regular
Gaussian profiles, the  4\,\AA\, full widths
as measured in two higher resolution spectra obtained in October,
and the absence of the high velocity component belonging to the accretion
stream, suggest an almost complete halt of the mass transfer.

The blue part of the continuum  completely changes
its shape when the system switches from the high to the low luminosity state.
The humps seen in the continuum during the high state become more pronounced,
and the blue end of the spectrum drops by  up to one order of magnitude.
In order to understand the composition of the system, we have to
examine the major contributors  of emission in polars.
In polars, except for photospheric emission from both components (which
can be readily excluded here), the widely accepted sources  of the flux
are the following four components originating
in the hot postshock region or its vicinity, where in-falling
matter above the magnetic pole slows down and heats  up before settling on the
surface of the WD:
\begin{itemize}
\item Hard X-ray bremsstahlung emitted from an optically
 thin column.
\item Blackbody radiation, thermalized and remitted by the
 heated surface of the WD in the soft X-ray and UV regions.
\item  Cyclotron emission from the column.
 It is emitted
 in a wide range of wavelengths from the near UV to the infrared region.
\item Recombination emission, depending on the strength
  of the illuminating X-rays.
\end{itemize}

We are primarily interested in the latter two as the  most probable sources
of light in the blue part of our spectra.

\subsection{Continuum Modelling}

\begin{figure}[b]
\includegraphics[width=80mm,bb=19 140 590 690]{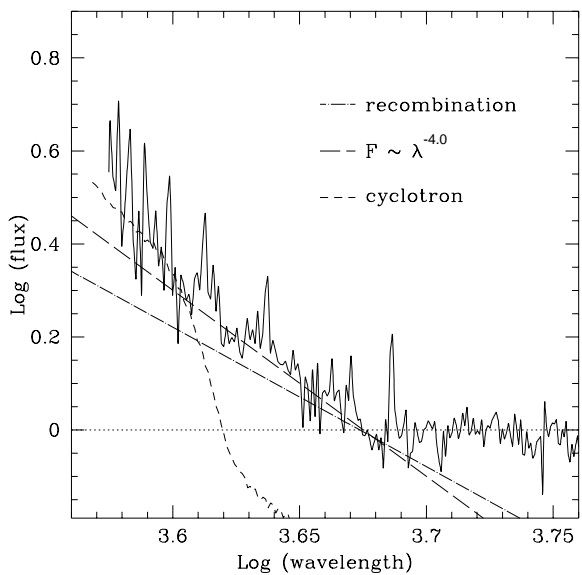}
\caption{The differential spectrum of \rxj\ between high and low luminosity states.
The additional emission responsible for changes is very ``hot''. The dashed line
representing power law is
drawn to give an idea of its steepness.
 Also plotted are flux distributions of
recombination emission and cyclotron optically thick  emission.
}
\label{diffsp}
\end{figure}

Examination of the spectral appearance in the low state, where
cyclotron lines are seen as broad humps in the continuum, and
comparison to identical objects (most remarkably HS\,1023+3900
(Reimers \cite{Reimers})), strongly advocate the cyclotron nature
of the observed flux. Actually, if not for tiny emission remaining
in the lower order Balmer lines, apparently coming from the
irradiated secondary, and significant contribution of secondary
photosphere in the continuum, the spectrum of \rxj\ in the low
state is very identical to that of HS\,1023+3900 during its low
state. The difference is only in the location of the cyclotron
humps due to the difference in magnetic field. Luckily, in the
case of \rxj\ we have observed the system  spectroscopically in
both luminosity states, high and low. In the high state, the
spectral appearance is notably different. We can see broader,
multicomponent emission lines, higher orders of Balmer lines and
appearance of not very strong He\,{\sc {ii}} emission in the
spectrum,  completely absent in the low state. In  addition, a hot
component changes the slop of the continuum in the blue side of
the spectrum. We subtracted the low state spectrum from the
averaged spectrum in the high state. The absence of orbital
coverage in the low state prevents us from studying phase related
effects. The resulting energy distribution, presented in Fig.
\ref{diffsp} clearly consists of two components. Redward of
4800\,\AA\ it is flat, while blueward it rises steeply (power law
index=-4\,!). It hardly can be attributed to the  blending by
Balmer lines in vicinity of the Balmer jump (the Balmer lines are
well resolved in the higher resolution spectra up to H$_{10}$, yet
continuum is consistent with the lower resolution spectra), nor to
the simple addition of a black body, e.g. from a hot spot. The
power law is presented by the dashed line in Fig. \ref{diffsp}, it
roughly indicates  a blackbody with T$_{eff}\gg$100\,000\,K.
%Neither the differential spectrum can be explained in terms of recombination
%emission

Alternatively to cyclotron emission, the low and high state
transition could potentially also be due to off/on states of
recombination emission. In the low state there is no He\,{\sc
{ii}} emission at all, and He\,{\sc i} is very weak, indicating
that the excitation by X-rays is very low. Thus, it is conceivable
that the  illumination of the stream is so low that also no
recombination emission is present. If, during the optical high
state, the accretion rate increases, X-ray emission sets in,
causing accordingly He\,{\sc {ii}} excitation, and also recombination
emission.
Normally in magnetic CVs one would expect recombination emission to
come  from stream where typical assumed temperatures are of the order of
20\,000\,K only.
 We calculated the  recombination emission energy
distribution for 20\,000\,K and also for hypothetical source
with temperatures as high as 5 and 15\,keV.
% using  the XSPEC package {\bf reference??}.
% which I usually
%use for X-ray data.
The input parameters are the edge energy (Paschen jump in our case),
the temperature and a normalization parameter.
%For hydrogen, the edge energies would be the
%Balmer and Paschen jumps (in our case probably only the Paschen jump),
%and the slope of the reco spectrum is computed by the model.
%
 The dot--dashed line in the
Fig. \ref{diffsp} corresponds to a 15\,keV temperature. Lower temperature
curves are naturally much flatter. Even 15\,keV source is apparently not
steep enough to satisfactory fit the additional source of
emission. The slope actually does not depend on the temperature of
the source, thus recombination emission solely can not be
responsible for explaining the transformation.

We sought an explanation of the continuum energy distribution of
\rxj\ and its changes in cyclotron emission, and found
satisfactory fits to explain both high and low luminosity states.
The solution is not commonly applicable to the rest of polars, but
we consider this object to be of different breed along with
HS\,1023+3900 due to its very low accretion rate, distinct from
polars outside the ``period gap''. There is  an extensive
theoretical  work on modelling the  transfer of radiation  through
the  standard accretion  column (\cite{Chanmugam, Meggit,
Barret1}). \cite{Barret2}  applied their models to fit the spectra
of the magnetic CV VV\,Pup.  We followed their procedures in
constraining our models in  order to describe the  observed flux
of  \rxj.  The authors demonstrate how the  absorption  and
emission  features of  cyclotron emission originate and how they
differ.  In order to obtain an absorption spectrum one has to
assume the presence of a cooler magnetized plasma layer with  a
hotter blackbody  (BB)  source  behind  it, representing  the
postshock region. On the other hand, an emission cyclotron
spectrum will arise from the  slab of  plasma representing  the
postshock  region  of the accretion column itself. The comparison
of sample models presented by \cite{Barret2}, demonstrates that in
the case of a pure  emission spectrum  the blue end of the
spectrum is rapidly decreasing. At the same time, the absorption
cyclotron spectrum, from an optically thick slab, is
Rayleigh-Jeans-like and exhibits a steep rise toward the blue,
similar to blackbody radiation.

Regarding the modelling  it is necessary to note that the
absorption cyclotron spectrum never was used in the actual
modelling of any  observed object, even by the authors of the
model. The above mentioned model does not take into account the
geometry of the magnetic spot, while there are modern models (e.g
Woelk \& Beuermann (\cite{wobe92}), Rousseau \etal (\cite{ro96})
and more recently Fischer \& Beuermann (\cite{fibe01})), that
successfully calculate corresponding parameters over the whole
white dwarf atmosphere. However, from one hand being restricted by
a few spectra in the low state not covering all orbital phases,
and on the other hand being able to achieve satisfactory results
in fitting the observed flux distribution by simple models, we did
not aim determination of exact parameters of the system by
applying state of the art models, rather than sought explanation
of its nature.

Thus, we used the  \cite{Barret2} model
to fit the  observed continuum of
\rxj\, by a least square  fit of
\begin{equation}
\sigma = \sqrt{\sum_{\lambda}{(F_{obs}(\lambda)-F_{mod}(\lambda))^2}}
\end{equation}
where $F_{obs}$ are relative fluxes normalized to  $F_{\lambda8100}=1$
in the low or high state with emission lines excluded from the analysis.
$F_{mod}$  is actually  a  sum
\begin{equation}
F_{mod} = a*F_{em}+b*F_{M4}+c*F_{abs},
\end{equation}
where  $F_{em}$ is   cyclotron emission from an optically thin medium,
while  $F_{abs}$ is cyclotron emission from optically thick
parts of the accretion column,  and  $F_{M4}$ is the spectrum of the
M4\,{\sc v} secondary star.
The coefficients  $a$, $b$ \& $c$     determine
the relative contributions  of the corresponding components.
$F_{em}$ and $F_{abs}$  are  functions
of the magnetic  field ($B$), the effective  temperature  of the  underlying
source ($kT$), the angle ($\Theta$) between  the magnetic field and the
line of sight,      and the dimensionless      parameter
$\Lambda=\omega_pL/\omega_Bc$. Here   $\omega_p$   is   the   plasma
frequency,  $\omega_B$  is the  cyclotron  frequency  and  $L$ is  the
thickness of the plasma slab.

\begin{figure*}[t]
\includegraphics[width=150mm]{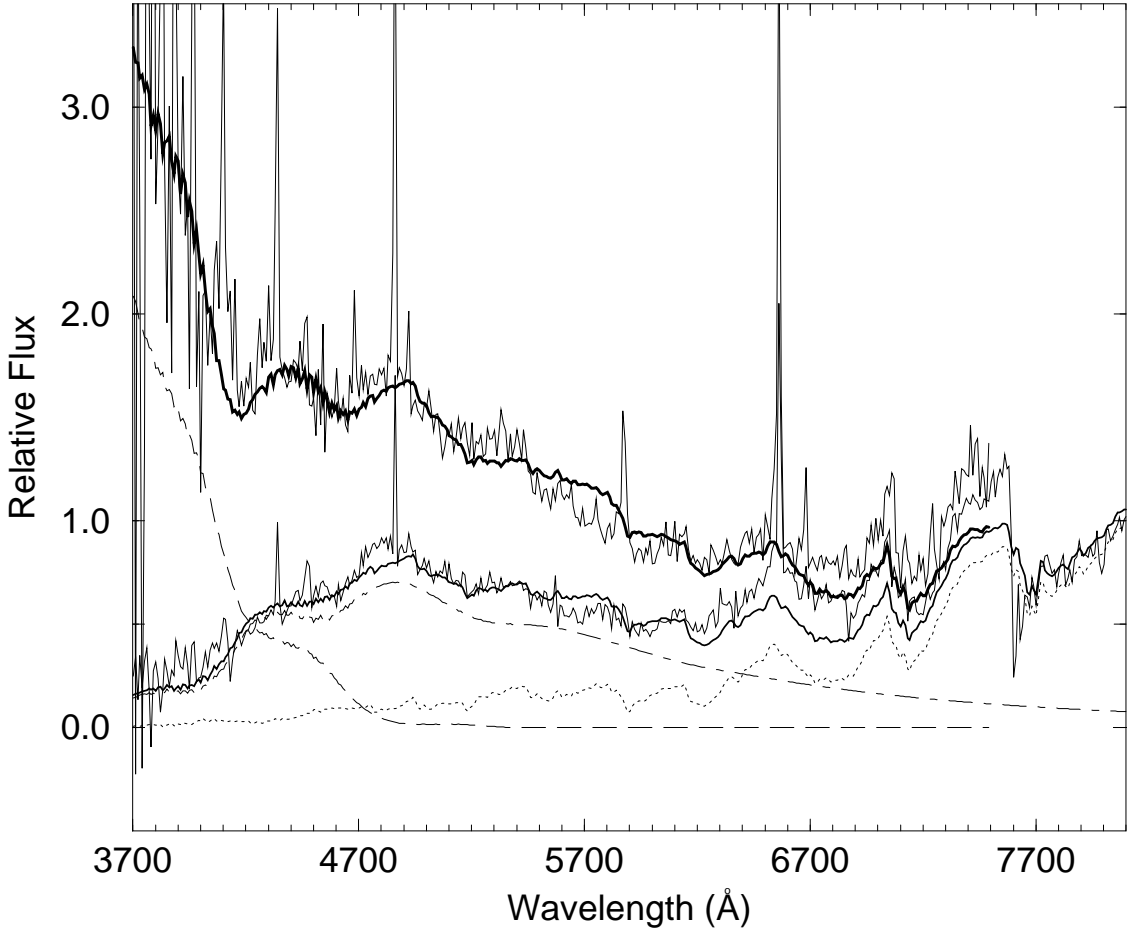}
\caption{ The low-resolution spectra of \rxj\ in high and low states are presented
   (thin solid line).
The dotted line corresponds to the M4\,{\sc v} star spectrum. Dashed
and dot-dashed lines are  the theoretical cyclotron absorption and emission
spectra of plasma in the magnetic field of the WD respectively. Thick solid
lines are re-presenting  the  resulting   fits to the  continuum of the \rxj\
in the  high and low states (see the discussion in 4.3).
}
\label{Photometry}
\end{figure*}

{\sl Low state spectrum:}~~
%We let all these parameters vary
%freely in order to reach the  best fit to  the low  state
%spectrum.
Usually in similar cases one assumes a temperature (say 10 or 15
or 20 keV), also assumes a $\Theta$  (say 70 degrees), in order to
find plausible parameters B and $\Lambda$,  which will place
synchrotron humps in the right places. Location of humps  strongly
depends on B and $\Lambda$ parameters, while temperature (kT) and
$\Theta$ parameter rather influence shape of the humps. Since in
most of the cases there are many other contributing components the
fitting becomes limited by the  ``cross-correlation'' of humps in
the  model with the  observed spectra. Location of humps becomes
crucial in determination of the magnetic field, which is the most
relevant parameter.

However in \rxj\, we have somehow a unique sort of low-state
spectrum, where in order to achieve  a good fit we do not need
anything else but secondary spectrum + $F_{em}$. Thus we can fit
$F_{em}$ much more precisely, than usually can be done. What we
did was to do least-squares along the spectrum with some initial
parameters floating them all four of them  and came up with the
result, very nicely describing the observed data.
%IT WAS NOT JUST "CROSSCORRELATION"
%of the humps in model and in observed spectrum.

The  hump at around H$\beta$ is very  distinct and it was easy to
find the corresponding initial  parameters, and  after wards
refine them with least-square fitting.   The contribution  of the
hot component $F_{abs}$ in the  low state  spectrum was  set to
zero (c=0). The calculated emission cyclotron spectrum is shown by
a thin dashed-dotted line in Fig. \ref{Photometry}. Its
combination with a M4\,{\sc v}, secondary star spectrum  (tiny
dotted line) gives a resulting best fit to the low-state observed
spectrum   shown as the lower thick solid line in Fig.
\ref{Photometry}. The following parameters were inferred from the
cyclotron emission model: $B=31$\,MG; $\Theta$ = 56\degr;
$kT=14$\,keV; $\Lambda = 6.6\times10^5$. Of course the shape of
spectrum should change with orbital phase, thus
 the obtained parameters will vary too. That basically refers to
$\Theta$ and $\Lambda$. Temperature is in range, which one would expect
to find in a polar, although it is higher than in HS\,1023+3900.
 The  M4\,{\sc v} secondary contribution is  $\sim$50\%
to the total flux in the R$_c$ filter and only $\sim$10\% in the B
band. The remaining deviation  of   the  modelled  spectrum   from
the observed  one can be explain as the consequence of the
deviation of the emission region  geometry from the simple and
homogeneous  geometry of the Barrett \& Chanmugam model,  and/or
the deviation of the radiation of  the optical thin plasma in the
magnetic field from thermal radiation.
\begin{figure}[t]
\includegraphics[width=130mm, bb=16mm 64mm 180mm 235mm, clip]{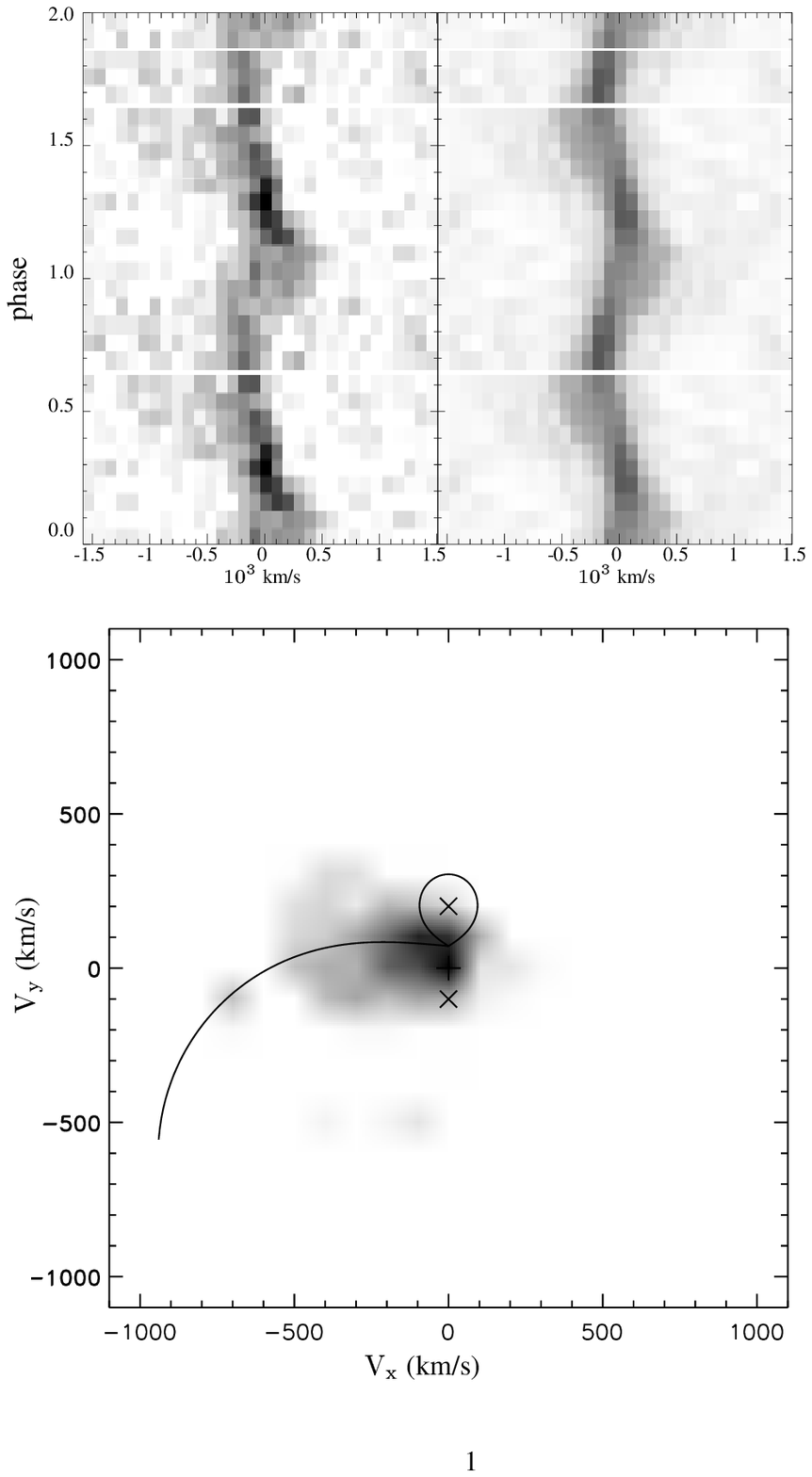}
\caption{Trailed,   continuum-subtracted,   spectra around He\,{\sc {ii}} of
 \rxj\, plotted in two cycles (upper panels).
Doppler map of  the He\,{\sc{ii}} emission line
 in  velocity space ($V_x,V_y$)  is given.  A  schematic overlay
marks the  Roche lobe of  the secondary and the ballistic  trajectory. The
secondary star and gas-stream trajectory are plotted for arbitrary mass ratio
q=0.5 (M$_{\rm 2}$=0.24\msun)  and inclination angle $i$=50\grad.
}
\label{doppler}
\end{figure}

{\sl  High state  spectrum:}~~ The examination  of the  spectra in
the high state demonstrate  that the   hump at around
$\lambda\,4800$\,\AA\ is still detected, although less pronounced.
Thus,  we believe that the spectrum in the  high-state is composed
of the same components as in the low state with the addition of  a
hot component  which we identify as cyclotron emission from the
optically thick portion
of the plasma slab in the model. This component is  responsible
for the  excess of  flux in the near UV region. It is much steeper
(in the wavelength range shortward of 4500\,\AA) than recombination
emission, as can be seen in Fig. \ref{diffsp}. Indeed, the
addition of an absorption cyclotron component shown as dashed line
in Fig. \ref{Photometry}, brings a good agreement between the
composite spectrum and the observed one in the high state. The
resulting fit to the  high state spectrum is shown as another
(upper) thick solid line in Fig. \ref{Photometry}. Assuming that
the  high luminosity is due to a higher accretion rate, we suppose
that an increase of matter in the accretion column may create
favorable conditions for the appearance of an optically thick,
absorption cyclotron spectrum. The parameters of the absorption
cyclotron emission which we obtained from the fitting confirm
this. With the magnetic field strength and angle remaining the same,
the parameter $\Lambda = 40.6\times10^5$ shows a five-fold
increase as compared to the emission spectrum from the optically
thin layer in the optical low state. This increase most probably
is conditioned by the increased density of the matter in the
column. Meanwhile, the temperature of the optically thick matter
fits better to the observed flux at $kT$= 18.5 keV, but it may
vary significantly without major impact on the final result. It is
worth mentioning, that in the high-state spectrum fits, both
cyclotron components contribute, showing that we have optically
thick and thin plasma at the same time. The temperature of
optically thin emission-cyclotron radiation however remains at
around 14\,keV as in the low state. But, the overall contribution
of the optically thin emission is increased in comparison to the
low state, which may be the result of an increase of the  emitting
area. The contribution of the corresponding components is
estimated as follows: In the red part of the spectrum (R$_c$ band)
the contribution of emission-cyclotron emission rises to 60\%,
with the rest coming from the secondary star. In the B band the
absorption cyclotron emission from the optically thick plasma
contributes up to $\sim$40\% of total flux, while the optically thin
plasma is responsible for $\sim$55\%, and the remaining $\sim$5\% are coming
from the secondary star. It is not excluded, however, that part of
the emission in the near UV has a recombination origin. In that
case the budget will be slightly different of course. From our
data it is difficult to estimate the relative contribution of
recombination emission and optically thick cyclotron absorption.
It is important, however, that recombination emission only will not be
enough to explain the blue/ultraviolet excess in the high state, and the
suggested scenario above may be a reasonable alternative.

\section{Doppler tomography}

Doppler  tomography \cite{Marsh}  is a  widely used
method to extract further information on CVs from trailed spectra.  In the
case  of magnetic CVs,  particularly polars,  Doppler tomography  in   most
cases reveals a very distinctive pattern:  the compact spot  appears at
the location  of the heated tip  of the secondary star  with a diffuse
arm extending from the L$_{\mathrm 1}$ point almost perpendicular at
first, then  bending toward negative  velocities (both V$_{\mathrm x}$
and   V$_{\mathrm  y}$).   It  follows   closely  the  calculated
trajectory of free falling matter from the Lagrangian point.

We  used the  code developed  by \cite{spruit}  to obtain a
Doppler  map of  \rxj\  using the maximum  entropy  method.  The  resulting
Doppler  map  (or  tomogram)  of  the He\,{\sc {ii}} emission  line  is
displayed  as  a  gray-scale  image  in  Fig. \ref{doppler}.   Also,
trailed spectra  of the He\,{\sc {ii}} line in
phase space  and its corresponding  reconstructed counterpart are displayed
in Fig. \ref{doppler}.  The
Roche lobe  of the  secondary calculated for   a
mass ratio  $q=0.5$ and an inclination angle $i$=50\grad\ is drawn on  the map.
These were selected from the values  estimated from the light curve
variations amplitude, due to the elliptically distortion of the secondary.
Also shown  is the accretion
stream   according   to    the   above   mentioned
parameters. The  resulting map  is another good  evidence in  favor of
the classification of \rxj\, as a polar.

\section{Conclusions}

We have studied  the new cataclysmic variable \rxj\, discovered by ROSAT
observations.
The  soft  X-ray spectrum  and  strong  X-ray variability  immediately
suggests  a  magnetic  cataclysmic  variable  nature.   The  follow-up
optical  spectral  and  photometric observations  provided  additional
support  for this  classification. The continuum spectrum shows  direct
evidence  of the presence  of a magnetic accreting  WD in  the system.
These, combined with  emission  line profiles, RV  curves and Doppler  maps,
and  characteristic  light curve  undoubtedly  supports the  magnetic
nature of \rxj.

We  derived P$_{\mathrm orb}$  = 2\fhour753  = 165.2  min for \rxj,
placing it inside  the period gap for CVs.  This  period is very close
(even shorter)  of the  recently discovered HS\,1023+3900 (Reimers \cite{Reimers}).
The latter  is considered  unique,  because it
exhibits a pure  cyclotron  emission  line spectrum.
Reimers (\cite{Reimers}) claims that  the appearance of the spectrum  is a result
of extremely low mass transfer rate, which in turn is supported by the
evolutionary scenarios, according to  which the mass transfer comes to
a halt in the period gap and systems become undetached.

\rxj\, is another system in this range of orbital periods showing a reduced
accretion rate, and actually barely any in the low state. The switches
between high and low states also demonstrate a little bit unusual
frequency, which might be the sign of a system at the verge of its accretion
era.
The continuous monitoring of this object may provide additional insights
into the behavior of the CVs and particularly magnetic ones in the period gap.

We calculated the cyclotron spectra for \rxj\, in the low and high states.
Our main conclusions on the origin of optical radiation in the continuum
are as follows:

1. \rxj\, has a  magnetic field strength of the WD of $3\times10^{7}$
 Gauss, and a postshock plasma temperature of 14 keV nicely consistent
 with the above assumed 20 keV hard X-ray component.

2. In the low state we have  pure emission cyclotron spectrum from
the optically thin cyclotron emitting source. In the high state we
have probably a higher accretion rate, and we assume that an
optically thick zone forms in the core of the accretion column.
Therefore, the high state spectra can be described by enhanced low
state emission plus additional radiation consisting of the
combination of an optically thick cyclotron-absorption spectrum
region and possibly recombination emission from reprocessed X-rays.

3.  The remaining small  deviations  between   the  modelled and
observed spectra of \rxj\, can be explained as a consequence of
the deviation of the emission region  geometry from the simple
geometry of the Barrett \& Chanmugam model  and/or the deviation
of the radiation of  the optical thin plasma in the magnetic field
from a thermal model.

We estimate the  mass and radius of the secondary star  as $M_2$  = 0.24\msun\
 and
$R_2 =0.29 R\sun$ from the  mass-period and radius-period
relations of \cite{Echevarria}. The mass is consistent
with the observed M4\,{\sc{v}} classification of the secondary and the
assumption that it is not much different from a main-sequence star of
corresponding temperature.

The mass ratio $q$ and inclination angle $i$ of the system were estimated
from the amplitude of ellipsoidal variations of the secondary in the low state
and independently confirmed and refined from Doppler tomography to be
around 0.5 and 50\grad,  respectively.

The distance to   \rxj\, also can be estimated from  the intensity
of the secondary in the spectra of \rxj\, in the low and high states.
The flux of the  M4\,{\sc{v}} star at 5500\,\AA\, or in the  V band,
is $6\times10^{-17}$  ${\mathrm {erg/cm^2/s/\AA}}$,
corresponding to a brightness of  $V=19.45$ magnitude
(for example, see   \cite{Fukugita}).
A M4{\sc{v}} star has  $M_V =12.95$ (for example  G\,165-8 in
\cite{Henry}).  Assuming a  maximum interstellar  absorption in
this direction  of $A_V=0.2\ \mathrm{mag}$ \cite{Dickey},
we estimate a distance to
this binary system of  about 180--200 pc.  Even if the high value of
the distance estimates were correct, the
correspondingly higher luminosity which we derive from the measured
flux would still be lower than  that of  most magnetic CVs.

With a distance estimate available,  we can also infer the accretion rate by
 assuming that the accretion
luminosity is emitted mostly in X-rays. We apply equation 6.10 from  Warner
\cite{Warner}  and  estimate $\dot{M}$ using the bolometric luminosity
from X-ray observation as $\dot{M} \sim
4\times10^{-12}$(d/180)$^2$ \msun\ yr$^{-1}$.

In this brief report we did not intent to derive all
parameters  of   this  very   interesting  new discovered polar.  New
observations are necessary to  study  this system in more detail. Meanwhile,
its existence seems to provide a good test bed for existing
evolutionary models.

\acknowledgements

We are grateful to A. Pramsky and A. Ugryumov for help in SAO RAS 6m
telescope observations.
This work was supported in part by CONACYT project 25454-E and a DGAPA
project.

\clearpage

\end{document}